# Study of the Effects of Artificial Dissipation and Other Numerical Parameters on Shock Wave Resolution

Frederico Bolsoni Oliveira[1*] and João Luiz F. Azevedo[2]

[1*]Instituto Tecnológico de Aeronáutica, São José dos Campos, 12228-900, São Paulo, Brazil.
[2]Instituto de Aeronáutica e Espaço, São José dos Campos, 12228-904, São Paulo, Brazil.

*Corresponding author(s). E-mail(s): fredericobolsoni@gmail.com;
Contributing authors: joaoluiz.azevedo@gmail.com;

**Abstract**

The effects induced by numerical schemes and mesh geometry on the solution of two-dimensional supersonic inviscid flows are investigated in the context of the compressible Euler equations. Five different finite-difference schemes are considered: the Beam and Warming implicit approximate factorization algorithm, the original Steger and Warming flux vector splitting algorithm, the van Leer approach on performing the flux vector splitting and two different novel finite-difference interpretations of the Liou AUSM$^+$ scheme. Special focus is given to the shock wave resolution capabilities of each scheme for the solution of an external supersonic inviscid flows around a blunt body. Significant changes in the shock structure are observed, mainly due to special properties of the scheme in use and the influence of the domain transformation procedure. Perturbations in the supersonic flow upstream of the shock are also seen in the solution, which is a non-physical behavior. Freestream subtraction, flux limiting and the explicit addition of artificial dissipation are employed in order to circumvent these problems. One of the AUSM$^+$ formulations presented here is seen to be particularly more robust in avoiding the appearance of some of these numerically-induced disturbances and non-physical characteristics in the solution. Good agreement is achieved with both numerical and experimental results available in the literature.

**Keywords:** Artificial Dissipation, Shock Wave Resolution, Blunt Body Flow, Bow Shock, Flux Vector Splitting Schemes

# 1 Introduction

In supersonic aerospace applications, the correct determination of shock wave behavior is of vital importance for the design of high-speed aircraft and spacecraft. One special case of such type of flow is the strong detached shock wave, known as a "bow shock", that develops around bluff bodies when immersed in supersonic and hypersonic flowfields. Although sharp leading edge bodies are capable of inducing the fluid to assume a bow shock configuration provided the correct Mach number is in effect, the present paper is mainly interest in bow shocks that arise from the interaction between the fluid and the solid walls of a blunt body. Such type of geometry is mainly found on the design of heat shields and nose cones of reentry vehicles. In such cases, the development of a bow shock can be seen as a desired effect, since it elevates the drag coefficient of the body which, in turn, decreases its descent velocity at the cost of a considerable increase on the average wall temperature. This flow condition has been the subject of intense research by the Computational Aerodynamics Laboratory of Instituto de Aeronáutica e Espaço (IAE) for the past few years [1, 2]. The group makes use of numerical tools developed in-house in order to deal with flows of this nature. For that reason, there has been a natural increase in interest in revisiting well-known, simple, test cases that enable the occurrence of strong shock waves in order to assess the shock capturing capabilities of the numerical formulations in use. This, in turn, provides the opportunity for a better understanding regarding the underlying effects of artificial dissipation over the predicted shock wave structure.

The accurate numerical determination of a shock structure can be heavily dependent on the numerical scheme used in the discretization of the model equations, as well as on the geometrical properties of the mesh. The present paper investigates these relations by applying five different numerical schemes, in the context of finite differences, for the discretization of the compressible 2-D Euler equations written in general curvilinear coordinates [3, 4]. This level of formulation is considered to be a reasonable approximation for the behavior of the supersonic flows of interest here. The chosen test case is a simple circular 2-D blunt body immersed in a supersonic steady flow, as described by Peery and Imlay [5]. For this case, the mesh was purposely designed with the intention of achieving a high curvature along the domain, thus enabling the observation of numerical problems that arise from such a condition. In order to have a basis for comparison, the simulation results are analyzed in relation to the work of Peery and Imlay [5], Lin [6] and Holden et al. [7].

The numerical solution of the proposed test case is obtained using an in-house CFD code that implements each one of these numerical methods [8]. The first scheme used is the implicit Beam and Warming approximate factorization scheme [9], which uses an implicit Euler time march coupled with a second-order centered scheme for the discretization of the spatial derivatives. The implicit linear system is solved using an alternating direction implicit (ADI) approximate factorization approach, which reduces the implicit matrix operator to a block-tridiagonal form and, thus, decreases the computational cost of solving the linear system exactly. Since this is a centered scheme, artificial dissipation must be added to the discretized equations in order to keep the numerical system within stability bounds. Furthermore, it is of interest that the addition of artificial dissipation does not introduce undesirable oscillations to the surroundings of the shock wave. Therefore, Pulliam's nonliner artificial dissipation model [10] is used in the explicit part of the equations, while a simplified version of it is applied to the implicit matrix operator.

The other four numerical schemes follow the flux vector splitting concept. The first one is the Steger and Warming original flux vector splitting scheme [11], coupled with an implicit Euler time march. Since this scheme can produce numerical oscillations near regions where the eigenvalues of the flux vector Jacobian matrices change signs [11], the van Leer approach for performing the flux vector splitting [12] is also employed here. Both of these upwind schemes are implemented using first and second order finite difference operators in order to observe the dissipative nature of lowering the order of the scheme. The other two schemes are different reinterpretations of Liou's Advection Upstream Splitting Method Plus (AUSM$^+$) explicit method [13], adapted to the context of finite differences. For these schemes, only a first order spacial discretization is considered. Since the original scheme was formulated using a finite volume framework, special care has to be taken when reinterpreting it using finite difference constructs. Hence, the authors propose two possible reinterpretations of the AUSM$^+$ scheme, both written in general curvilinear coordinates.

To summarize, the main purpose of the present effort is to address the relevant numerical issues that might influence the accurate calculation of external aerospace flows in the presence of strong shock waves. The authors are mainly concerned with the capability of each scheme to correctly capture the freestream properties without numerical disturbances and to preserve the shock wave profile. The influence that techniques such as freestream subtraction and explicit addition of artificial dissipation have on the predicted shock wave location is also considered. Furthermore, the numerical properties of the proposed AUSM$^+$ formulations are also assessed using the aforementioned test case.

## 2 Two-dimensional formulation

The formulation to be presented in this section has been the subject of extensive study by the authors' research group throughout the years, both in its two-dimensional and three-dimensional forms. It has been applied to the simulation of high-fidelity jet flows [3, 4], internal duct flows [8, 14], external transonic and supersonic flows [15–18], among many others. The aim of the present section is to briefly introduce the two-dimensional form of the equations used to model the fluid flow of interest. Therefore, we refer the interested reader to the previously mentioned references for more details about the development and usage of the current theoretical formulation.

The two-dimensional Euler equations can be written in conservation-law form and curvilinear coordinates as follows [8, 19, 20]:

$$\frac{\partial \hat{Q}}{\partial \tau} + \frac{\partial \hat{E}}{\partial \xi} + \frac{\partial \hat{F}}{\partial \hat{\eta}} = 0 \, , \tag{1}$$

where $\hat{Q}$ is the vector of conserved variables:

$$\hat{Q} = J^{-1} \begin{Bmatrix} \rho \\ \rho u \\ \rho v \\ e \end{Bmatrix} \, , \tag{2}$$

and $\hat{E}$ and $\hat{F}$ are the flux vectors for the inviscid case:

$$\hat{E} = J^{-1} \begin{Bmatrix} \rho U \\ \rho u U + \xi_x p \\ \rho v U + \xi_y p \\ U(e+p) - \xi_t p \end{Bmatrix} \, , \qquad \hat{F} = J^{-1} \begin{Bmatrix} \rho V \\ \rho u V + \eta_x p \\ \rho v V + \eta_y p \\ V(e+p) - \eta_t p \end{Bmatrix} \, . \tag{3}$$

Here, $\rho$ is the local fluid density, $u$ and $v$ are the Cartesian components of the velocity vector, $e$ is the total energy per unity of volume and $p$ is the static pressure. Variables $\xi$ and $\eta$ are the space coordinates of the curvilinear domain, $\tau$ is the time variable in the new system (made equal to $t$), and $J$ is the Jacobian of the domain transformation, expressed as:

$$J^{-1} = x_\xi y_\eta - x_\eta y_\xi \, , \tag{4}$$

where the subscript means a partial derivative operator. The mapping from the Cartesian domain to the transformed domain is made in such a way that only discrete unitary variations occur between two adjacent points. The metric terms $\xi_t$, $\xi_x$, $\xi_y$, $\eta_t$, $\eta_x$ and $\eta_y$ are calculated from the derivatives of the Cartesian coordinates as shown below:

$$\xi_x = J y_\eta \, , \qquad \xi_y = -J x_\eta \, , \qquad \xi_t = -x_\tau \xi_x - y_\tau \xi_y \, , \tag{5}$$
$$\eta_x = -J y_\eta \, , \qquad \eta_y = J x_\xi \, , \qquad \eta_t = -x_\tau \eta_x - y_\tau \eta_y \, . \tag{6}$$

The $U$ and $V$ variables are the contravariant components of the velocity vector. They can be expressed as functions of the metric terms as follows:

$$U = \xi_t + \xi_x u + \xi_y v \, , \qquad V = \eta_t + \eta_x u + \eta_y v \, . \tag{7}$$

For a calorically perfect gas, the pressure $p$ can be expressed as:

$$p = (\gamma_{gas} - 1) \left[ e - \frac{1}{2} \rho (u^2 + v^2) \right] \, , \tag{8}$$

where $\gamma_{gas}$ is the ratio of specific heats and, in the present application, is constant and equal to 1.4. All variables mentioned so far are used, during the computational loop, in their nondimensional form by taking suitable reference values.

# 3 Numerical schemes

## 3.1 Centered scheme

All of the numerical schemes employed here are written in the context of finite differences. Four different schemes are considered. The first one is Beam and Warming's implicit approximate factorization scheme [9]. For each node of the discrete domain, this scheme can be written as [8]

$$\left(I + \Delta t \delta_\xi \hat{A}^n + D_{\xi_{impl.}}\right)\left(I + \Delta t \delta_\eta \hat{B}^n + D_{\eta_{impl.}}\right)\Delta \hat{Q}^n = -\Delta t \left[\delta_\xi \hat{E}^n + \delta_\eta \hat{F}^n\right] + D_\xi + D_\eta \qquad (9)$$

where $I$ is the identity matrix, $\Delta t$ is the value of the time-step to be performed and $\hat{A}$ and $\hat{B}$ are the Jacobian matrices of the flux vectors $\hat{E}$ and $\hat{F}$, respectively [8]. Moreover, $\delta_\xi$ and $\delta_\eta$ are second-order central finite difference operators in the $\xi$ and $\eta$ directions, respectively. If $(i,j)$ are the two-dimensional indexes of a given node in a structured mesh, then the second-order central finite difference operators are defined as

$$\begin{aligned}\delta_\xi(\,\cdot\,)_{i,j} &\equiv \frac{(\,\cdot\,)_{i+1,j} - (\,\cdot\,)_{i-1,j}}{2} \;, \\ \delta_\eta(\,\cdot\,)_{i,j} &\equiv \frac{(\,\cdot\,)_{i,j+1} - (\,\cdot\,)_{i,j-1}}{2} \;, \end{aligned} \qquad (10)$$

considering a transformed domain with unit spacing. Since this is a centered scheme, it requires the addition of explicit artificial dissipation in order to damp high frequency oscillations that might cause the algorithm to diverge. In the present work, Pulliam's nonlinear artificial dissipation model [10] is used in the explicit component of the equations, while a simplified version of it is used in the implicit operator. In order to lay down the foundation of the discussion that takes place in Section 5, it is important to present the numerical formulation of this model. The explicit components of the Pulliam nonlinear artificial dissipation model [10], $D_\xi$ and $D_\eta$, are constructed using 2nd and 4th differences and have the following form:

$$\begin{aligned}D_\xi &= \nabla_\xi \left(\frac{\sigma^n_{i+1,j}}{J_{i+1,j}} + \frac{\sigma^n_{i,j}}{J_{i,j}}\right)\left(\epsilon^{(2)}_{i,j}\Delta_\xi \hat{Q}^n_{i,j} - \epsilon^{(4)}_{i,j}\Delta_\xi \nabla_\xi \Delta_\xi \hat{Q}^n_{i,j}\right) \;, \\ D_\eta &= \nabla_\eta \left(\frac{\sigma^n_{i,j+1}}{J_{i,j+1}} + \frac{\sigma^n_{i,j}}{J_{i,j}}\right)\left(\epsilon^{(2)}_{i,j}\Delta_\eta \hat{Q}^n_{i,j} - \epsilon^{(4)}_{i,j}\Delta_\eta \nabla_\eta \Delta_\eta \hat{Q}^n_{i,j}\right) \;.\end{aligned} \qquad (11)$$

In Eq. (11), the operators $\Delta_\xi$ and $\nabla_\xi$ are first-order forward and backward difference operators in the $\xi$ direction, respectively. These one-sided operators are defined in Eq. (12) considering a transformed domain with unit spacing.

$$\begin{aligned}\Delta_\xi(\,\cdot\,)_{i,j} &\equiv (\,\cdot\,)_{i+1,j} - (\,\cdot\,)_{i,j} \\ \nabla_\xi(\,\cdot\,)_{i,j} &\equiv (\,\cdot\,)_{i,j} - (\,\cdot\,)_{i-1,j}\end{aligned} \qquad (12)$$

The way in which the $\epsilon$'s are defined is what makes the model behave in a nonlinear manner. The expressions are:

$$\begin{aligned}\epsilon^{(2)}_{i,j} &= K_2\,\Delta t\,\max\left(\gamma_{i+1,j},\gamma_{i,j},\gamma_{i-1,j}\right)\;, \\ \epsilon^{(4)}_{i,j} &= \max\left(0, K_4\Delta t - \epsilon^{(2)}_{i,j}\right)\;,\end{aligned} \qquad (13)$$

where typical values of the constants are $K_2 = 1/4$ and $K_4 = 1/100$. However, in the present authors' experience, values as low as $K_2 = 1/8$ and $K_4 = 1/200$ have still been capable of keeping the system stable provided that the mesh is fine enough. The $\gamma$ function is a pressure gradient "sensor", used for reducing the original 4th difference model to a 2nd difference one in the neighborhood of pressure discontinuities (i.e., shockwaves). This procedure is done in order to damp the oscillations intrinsic to the behavior of the central 4th difference scheme near these regions. Together with the sum, $\sigma$, of the spectral radii of $\hat{A}$ and $\hat{B}$, the following function definitions are used:

$$\gamma_{i,j} = \frac{|p_{i+1,j} - 2p_{i,j} + p_{i-1,j}|}{|p_{i+1,j} + 2p_{i,j} + p_{i-1,j}|},$$
$$\sigma_{i,j} = |U| + a\sqrt{\xi_x^2 + \xi_y^2} + |V| + a\sqrt{\eta_x^2 + \eta_y^2}, \tag{14}$$

where $a$ is the local speed of sound. Equivalent expressions for Eq. (11), Eq. (13) and Eq. (14) are used for the $\eta$ direction. For the implicit operators $D_{\xi_{impl.}}$ and $D_{\eta_{impl.}}$, a simple 2nd difference operator is used. This is done in order to keep the overall structure of the implicit operator as two-sets of block tridiagonal matrices. This linear system structure is cheaper to compute than a block pentadiagonal system of equations, which would be achieved by using a 5-point, 4th-difference, central implicit operator. Thus

$$D_{\xi_{impl.}} = -\epsilon_{impl.} J^{-1} \nabla_\xi \Delta_\xi J, \qquad \epsilon_{impl.} = 3\left(\epsilon^{(2)} + \epsilon^{(4)}\right). \tag{15}$$

Similar definitions can be made for the $\eta$ direction.

## 3.2 Upwind schemes

The other four numerical schemes used here are upwind methods that follow the flux vector splitting approach. The first one is the Steger and Warming scheme [11], coupled with an implicit Euler time march. The second method is the van Leer scheme [12], which has the same overall structure as the Steger and Warming scheme but with different definitions for the split flux vectors. Essentially, this method keeps the eigenvalues of the Jacobian matrices of each split flux vector continuously differentiable throughout all values of local Mach number, which is not true for the Steger and Warming scheme. This behavior improves the robustness of the method, as well as the quality of the solution in regions close to discontinuities. Both schemes are written using first and second-order, one-sided, finite difference operators. The authors refer to the original papers [11, 12] for more details regarding the numerical formulation of these two schemes.

The last scheme is Liou's sequel to the Advection Upstream Splitting Method, referred to as AUSM$^+$ [13, 21], which follows a distinct approach for performing the splitting of the flux vectors. The fundamental idea behind it is to interpret the flux vectors as being composed of two different parts, one is responsible for the advection of passive scalars through the velocity field and the other is the transport of pressure terms by means of a pressure flux. These two fluxes, which are interpreted as numerical fluxes, are split by employing different strategies. Furthermore, a special definition for the reconstruction of the speed of sound evaluated at a numerical interface between two adjacent nodes of the domain is used. This approach is performed in order to keep the unification between the Mach number and the corresponding local velocity.

In its original form, the AUSM$^+$ scheme is written using a finite volume formulation [13, 21]. However, in order to be able to use the programming framework that was created by the authors during the implementation of the other three methods, the AUSM$^+$ scheme is reinterpreted in the context of finite differences and general curvilinear coordinates. Since the resulting formulation differs from the original, it is important to present it here. First, it is necessary to define the speed of sound, $a$, evaluated at the interface $i+1/2$, which is taken to be located at the midpoint between two adjacent nodes of the transformed domain. Considering $\kappa$ as being either the $\xi$ or $\eta$ direction, $a_{i+\frac{1}{2}}$ can be calculated as:

$$a_{i+\frac{1}{2}} = min\left(\tilde{a}_i, \tilde{a}_{i+1}\right), \qquad \tilde{a} = \frac{a^{*^2}}{\max\left(a^*, \left|\frac{\kappa_x u + \kappa_y v}{\sqrt{\kappa_x^2 + \kappa_y^2}}\right|\right)}, \tag{16}$$

where $a^*$ is the local critical speed of sound. Then, two functions $\mathcal{M}$ and $\mathcal{P}$ are defined in order to perform the split of the Mach number and the pressure terms. Following the same rationale behind the van Leer scheme, these functions ensure that the derivatives of the reconstructed fluxes are continuous across all flow regimes. The $\mathcal{M}$ function is defined as follows:

$$\mathcal{M}^{\pm}(\tilde{M}) = \begin{cases} \frac{1}{2}(\tilde{M} \pm |\tilde{M}|), & \text{if } |\tilde{M}| \geq 1, \\ \mathcal{M}_{\beta}^{\pm}(\tilde{M}), & \text{otherwise.} \end{cases} \qquad (17)$$

$$\mathcal{M}_{\beta}^{\pm}(\tilde{M}) = \pm \frac{1}{4}(\tilde{M} \pm 1)^2 \pm \beta(\tilde{M}^2 - 1)^2, \qquad -\frac{1}{16} \leq \beta \leq \frac{1}{2},$$

where $\beta$ is a control parameter whose recommended [13, 21] value is $\beta = 1/8$. For the $\mathcal{P}$ function, we have:

$$\mathcal{P}^{\pm}(\tilde{M}) = \begin{cases} \frac{1}{2}\left(1 + sign(\tilde{M})\right), & \text{if } |\tilde{M}| \geq 1, \\ \mathcal{P}_{\alpha}^{\pm}(\tilde{M}), & \text{otherwise.} \end{cases} \qquad (18)$$

$$\mathcal{P}_{\alpha}^{\pm}(\tilde{M}) = \frac{1}{4}(\tilde{M} \pm 1)^2 (2 \mp \tilde{M}) \pm \alpha \tilde{M}(\tilde{M}^2 - 1)^2, \qquad -\frac{3}{4} \leq \alpha \leq \frac{3}{16}.$$

Once again, $\alpha$ is a control parameter whose recommended [13, 21] value is $\alpha = 3/16$. The values of $\tilde{M}$ present in both Eq. (17) and Eq. (18) are local Mach numbers in the $\kappa$ direction evaluated using $a_{i+1/2}$:

$$\tilde{M}_i = \frac{1}{a_{i+\frac{1}{2}}} \left( \frac{\kappa_x u + \kappa_y v}{\sqrt{\kappa_x^2 + \kappa_y^2}} \right)_i, \qquad \tilde{M}_{i+1} = \frac{1}{a_{i+\frac{1}{2}}} \left( \frac{\kappa_x u + \kappa_y v}{\sqrt{\kappa_x^2 + \kappa_y^2}} \right)_{i+1}. \qquad (19)$$

The Mach number $m_{i+1/2}$ evaluated at the interface $i + 1/2$ can, then, be defined as:

$$m_{i+\frac{1}{2}} = \mathcal{M}^{+}(\tilde{M}_i) + \mathcal{M}^{-}(\tilde{M}_{i+1}). \qquad (20)$$

The split of this value can be achieved in a similar manner to the split of eigenvalues used in the Steger and Warming method:

$$m_{i+\frac{1}{2}}^{\pm} = \frac{m_{i+\frac{1}{2}} \pm \left| m_{i+\frac{1}{2}} \right|}{2}. \qquad (21)$$

The pressure component, $p$, evaluated at the numerical interface $i + 1/2$ follows the same idea:

$$p_{i+\frac{1}{2}} = \mathcal{P}^{+}(\tilde{M}_i) p_i + \mathcal{P}^{-}(\tilde{M}_{i+1}) p_{i+1}. \qquad (22)$$

In order to define a suitable numerical flux vector at the $i + 1/2$ interface, it is necessary to reconstruct the metric terms and the Jacobian that accompany the pressure variable in Eq. (3). Two different approaches are considered. The first one uses the same $\mathcal{P}$ function, previously employed in Eq.(22), in order to reconstruct the unknown terms. Hence, the generic numerical flux vector, $f_{i+1/2}$, is computed as:

$$f_{i+\frac{1}{2}} = a_{i+\frac{1}{2}} \left[ m_{i+\frac{1}{2}}^{+} \left( \frac{\sqrt{\kappa_x^2 + \kappa_y^2}}{J} \right)_i \begin{Bmatrix} \rho \\ \rho u \\ \rho v \\ e + p \end{Bmatrix}_i + m_{i+\frac{1}{2}}^{-} \left( \frac{\sqrt{\kappa_x^2 + \kappa_y^2}}{J} \right)_{i+1} \begin{Bmatrix} \rho \\ \rho u \\ \rho v \\ e + p \end{Bmatrix}_{i+1} \right] + \\ + \begin{Bmatrix} 0 \\ \mathcal{P}^{+}(\tilde{M}_i) \left( \frac{p \kappa_x}{J} \right)_i + \mathcal{P}^{-}(\tilde{M}_{i+1}) \left( \frac{p \kappa_x}{J} \right)_{i+1} \\ \mathcal{P}^{+}(\tilde{M}_i) \left( \frac{p \kappa_y}{J} \right)_i + \mathcal{P}^{-}(\tilde{M}_{i+1}) \left( \frac{p \kappa_y}{J} \right)_{i+1} \\ 0 \end{Bmatrix}. \qquad (23)$$

This numerical flux vector can either be equal to $\hat{E}_{i+\frac{1}{2},j}$ or $\hat{F}_{i,j+\frac{1}{2}}$, depending on whether $\kappa$ is taken to be $\xi$ or $\eta$, respectively. It should be noted, however, that by multiplying the metric terms by $\mathcal{P}^{\pm}$ effectively applies the same operator bias used in the upwinding of the pressure scalars to terms that are solely related to the geometry of the mesh. Hence, a second approach is also considered. Now, the metric terms, as well as the Jacobian, are reconstructed by performing a simple average of the adjacent nodal values. Therefore, the following formulation is obtained:

$$(\kappa_x)_{i+\frac{1}{2}} \equiv \frac{(\kappa_x)_i + (\kappa_x)_{i+1}}{2} \qquad (\kappa_y)_{i+\frac{1}{2}} \equiv \frac{(\kappa_y)_i + (\kappa_y)_{i+1}}{2} \qquad (J)_{i+\frac{1}{2}} \equiv \frac{(J)_i + (J)_{i+1}}{2} \quad (24)$$

$$f_{i+\frac{1}{2}} = \left(\frac{a\sqrt{\kappa_x^2 + \kappa_y^2}}{J}\right)_{i+\frac{1}{2}} \left[m^+_{i+\frac{1}{2}} \begin{Bmatrix} \rho \\ \rho u \\ \rho v \\ e+p \end{Bmatrix}_i + m^-_{i+\frac{1}{2}} \begin{Bmatrix} \rho \\ \rho u \\ \rho v \\ e+p \end{Bmatrix}_{i+1} \right] + \begin{Bmatrix} 0 \\ \left(\frac{p\,\kappa_x}{J}\right)_{i+\frac{1}{2}} \\ \left(\frac{p\,\kappa_y}{J}\right)_{i+\frac{1}{2}} \\ 0 \end{Bmatrix} \quad (25)$$

where the interface pressure and metric terms are calculated using Eqs. (22) and (24), respectively. Throughout this work, the first and second approaches are referred to as "Ap.1" and "Ap.2", respectively. A time march can be performed through the usual formula for an explicit scheme as:

$$\hat{Q}^{n+1} = \hat{Q}^n - \Delta t \left(\hat{E}^n_{i+\frac{1}{2},j} - \hat{E}^n_{i-\frac{1}{2},j} + \hat{F}^n_{i,j+\frac{1}{2}} - \hat{F}^n_{i,j-\frac{1}{2}}\right). \quad (26)$$

Due to the form in which the properties are extrapolated to the interface, the scheme given by Eq. (26) is only capable of achieving 1st-order spacial accuracy. Nevertheless, the scheme is considered to be suitable for performing the intended analysis.

### 3.3 Limiting procedure

As discussed in later sections, some numerical problems can arise from the usage of the unmodified version of these schemes in regions near strong discontinuities. Therefore, some modifications have been implemented in order to address such issues. The first one of them is to add a flux limiter to the 2nd-order upwind schemes. The approach adopted here is documented by Anderson et al. [22] and consists in introducing the limiting function directly to the definition of the one-sided 2nd-order finite difference operators. The 2nd-order, backward, finite difference operator, $\delta^-$, can be written in the following general form:

$$\delta^- b_i = \nabla b_i + \phi^+_{i-\frac{1}{2}} \frac{1}{2\Delta x}(b_i - b_{i-1}) - \phi^+_{i-\frac{3}{2}} \frac{1}{2\Delta x}(b_{i-1} - b_{i-2}), \quad (27)$$

in which $\phi$ is the limiting function and $\nabla$ is the 1st-order, backward, finite difference operator in physical space, $\nabla b_i \equiv (b_i - b_{i-1})/\Delta x$.

In Eq. (27), the 1st-order backward operator can be obtained by making $\phi = 0$, while the 2nd-order scheme is obtained by making $\phi = 1$. Second-order (and above) one-sided operators are known for introducing undesired numerical oscillations near discontinuities in the solution domain. Therefore, it is of interest to enable the one-sided operator to continuously decay to 1st-order whenever the local node is located near such condition. In the present work, a simple and suitable $\phi$ function is the minmod flux limiter [23]:

$$\phi^+_{i-\frac{1}{2}} = \max\left(0, \min\left(1, r^+_{i-\frac{1}{2}}\right)\right), \quad (28)$$

in which $r^+_{i-\frac{1}{2}}$ is the ratio of consecutive property variations:

$$r^+_{i-\frac{1}{2}} \equiv \frac{b_{i+1} - b_i}{b_i - b_{i-1}}. \quad (29)$$

Notice that, although all equations are shown for the backward one-sided operator, analogous definitions can also be made for the forward operator. These new definitions for the forward and backward operators are used here in the implementation of the 2nd-order Steger and Warming and van Leer schemes. Hence, the limiter function is acting upon the finite difference operator of the flux terms, constraining the slope of the 2nd-order corrections to the 1st-order scheme.

Another problem that must be accounted for is the fact that the discrete Euler equations written in conservation-law form and curvilinear coordinates, as shown in Eq. (1), do not "accept" the freestream state as a possible solution when applied to a non-uniform mesh. This same behavior is also true for the full Navier-Stokes system of equations. This is due to the presence of space-varying metric terms that are computed using finitely accurate finite difference schemes. In order to reduce the influence of this problem over the discrete solution, a freestream subtraction [24] is performed on the calculation of the fluxes of the Euler equations, as follows:

$$\frac{\partial \hat{Q}}{\partial \tau} + \frac{\partial \left(\hat{E} - \hat{E}_\infty\right)}{\partial \xi} + \frac{\partial \left(\hat{F} - \hat{F}_\infty\right)}{\partial \eta} = 0, \tag{30}$$

where the $\infty$ subscript refers to the constant freestream state. With this modification, it becomes clear that the freestream state is recovered as a possible discrete solution to the system of equations.

# 4 Description of test case and boundary conditions

The test case considered here is a simple circular blunt body exposed to a cold gas, supersonic freestream. The intention is to observe the ability of each scheme to correctly capture the location and geometry of the detached shock that develops in the upstream region of the body surface. The overall geometry of the blunt body consists of a circular solid body with a 76.2 mm diameter, $d$, immersed in a supersonic flow. The fluid of interest is taken to be atmospheric air with a constant ratio of specific heats, $\gamma$, of 1.4 and a gas constant, $R$, of 287 J/(kg K). The freestream boundary condition follows Peery and Imlay [5] and Holden et al. [7], in which the Mach number, $M$, static pressure, $p$, and total temperature, $T_t$, are set to 8.0, 855 Pa and 1726 K, respectively. These conditions are sufficient to constrain the subsonic region of the flow to a small fluid pocket located between the shock wave and the body surface, around the stagnation point. Therefore, the discrete domain can be limited to only a small region that surrounds the upstream part of the body, as long as the flow that crosses its boundaries is at full supersonic conditions. Hence, the subsonic wake that may develop downstream will not affect the results obtained for this isolated supersonic region. Moreover, since we are not concerned with the wake, the flow can be adequately described by the Euler equations. A schematic diagram of the problem setup as well as the expected physical behavior are shown in Fig. 1.

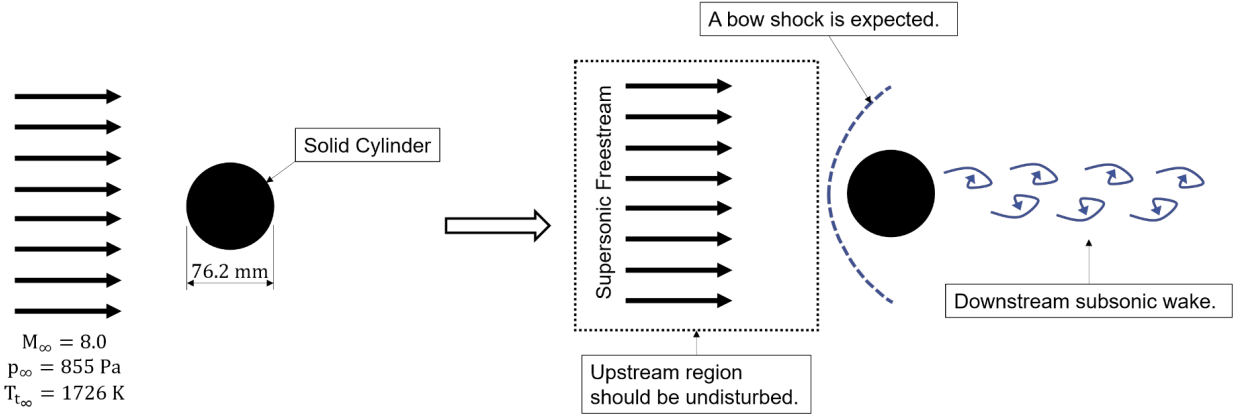

**Fig. 1** Schematic diagram of the blunt body problem considered here. The case setup is shown on the left, while the expected physical behavior is shown on the right.

The computational domain is defined as shown in Fig. 2, where the origin is located at the body geometrical center, with the $x$ axis pointing towards the right direction and the $y$ axis pointing upwards. The outer boundary is an arc of a circle with a diameter of $3.5d$. A structured mesh is, then, created using $100 \times 100$ nodes distributed with a slight clustering towards the wall region, as illustrated in Fig. 3. One should notice that the mesh lines are purposely created such that they do not follow the shock surface. The present code has been verified by the authors in a previous publication [8] and was seen to be capable of converging to a mesh independent solution provided a sufficiently refined mesh is used, as expected. The current mesh refinement was purposely chosen in order to facilitate the measurement and visualization of the numerical problems of interest.

As illustrated in Fig. 2, the right boundary is taken to be a solid wall, the left boundary is the supersonic freestream itself and the upper and lower boundaries are supersonic outflow regions. The properties along these outflow boundaries are calculated using zeroth-order extrapolation. Here, the solution is considered to be converged when the $L_\infty$ norm of the residue associated with the continuity equation decreases by 11 orders of magnitude or more, unless otherwise stated.

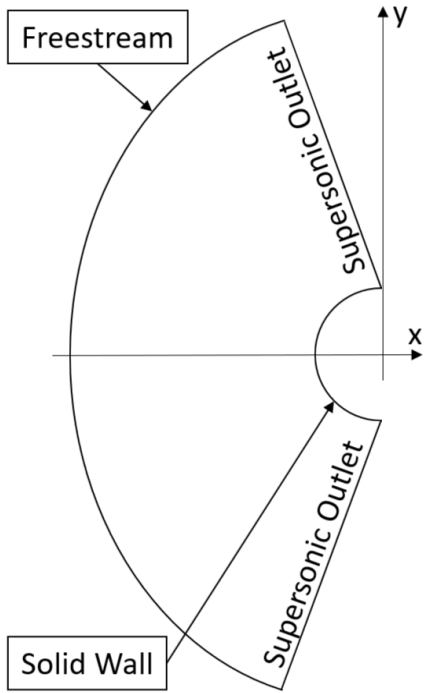

**Fig. 2** Schematic diagram of the boundary conditions applied to the blunt body case.

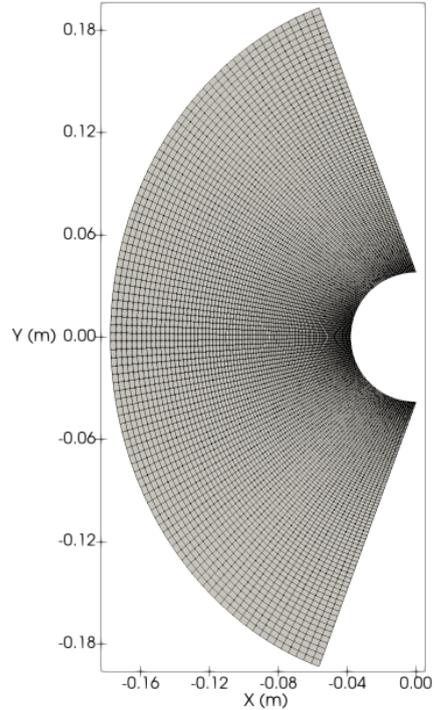

**Fig. 3** View of the computational domain defined for the simulation as well as the structured mesh used ($100 \times 100$ nodes). Coordinate units are in meters.

## 5 Results and discussion

Contour plots of the converged Mach number distribution in the steady state flow are shown in Figs. 4, 5 and 6 for the first-order Steger and Warming, van Leer and $\text{AUSM}^+_{Ap.1}$ upwind schemes, respectively, without performing the freestream subtraction. As one can see, although all three schemes were able to achieve similar results in terms of the macro properties of the flow, such as overall bow shock location and Mach number spatial distribution, there are a few inconsistencies of numerical nature that surround the domain symmetry line ($x$ axis), in the region upstream of the shock. In all three cases, there are nonphysical oscillations that appear in the upstream region of the bow shock, where an undisturbed freestream state is expected. This numerical problem is more apparent in the case of the first-order Steger and Warming scheme, as one can see in Fig. 4.

A better view of these numerical oscillations can be achieved if a flow property, such as the Mach number, is plotted along a vertical line located in the upstream region of the shock. The chosen vertical line is located at $x = -0.07\,m$, and goes from $y = -0.03\,m$ to $y = +0.03\,m$. Mach number values obtained for each one of the three first-order schemes considered so far are plotted along the vertical line in Fig. 7. For comparison, results obtained for the first-order $\text{AUSM}^+_{Ap.2}$ scheme are also plotted in the same figure. A considerable deviation from the analytic constant freestream state of $M_\infty = 8.0$ is observed in the results obtained with three of the schemes. The most critical case is seen in the first-order Steger and Warming scheme, whose

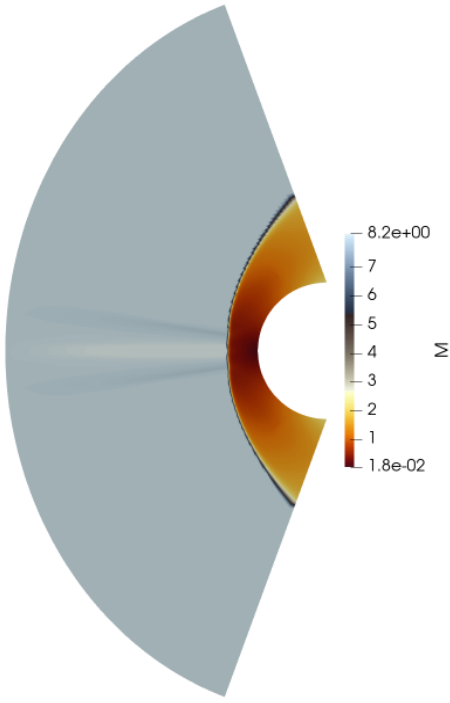

**Fig. 4** Calculated Mach number distribution for the inviscid blunt body flow using the first-order Steger and Warming upwind scheme without using freestream subtraction.

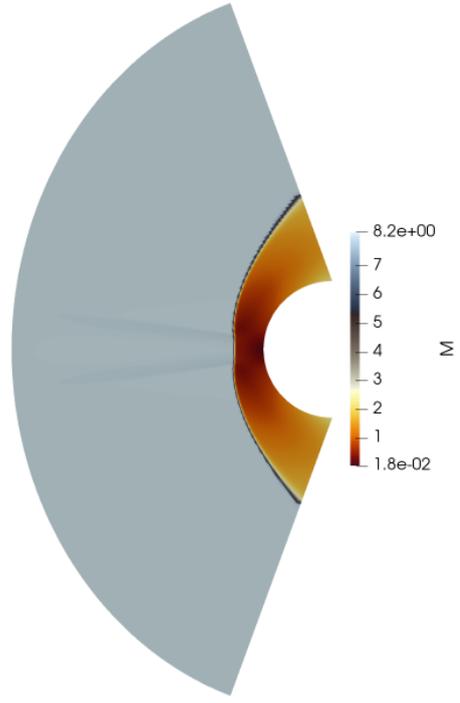

**Fig. 5** Calculated Mach number distribution for the inviscid blunt body flow using the first-order van Leer upwind scheme without using freestream subtraction.

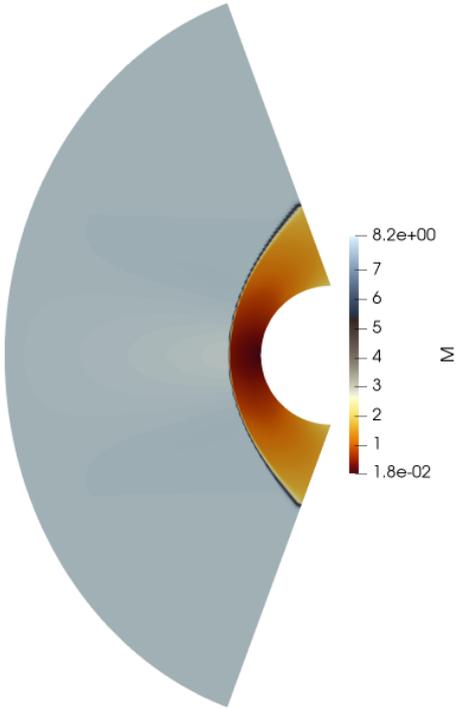

**Fig. 6** Calculated Mach number distribution for the inviscid blunt body flow using the first-order $\text{AUSM}^+_{Ap.1}$ upwind scheme without using freestream subtraction.

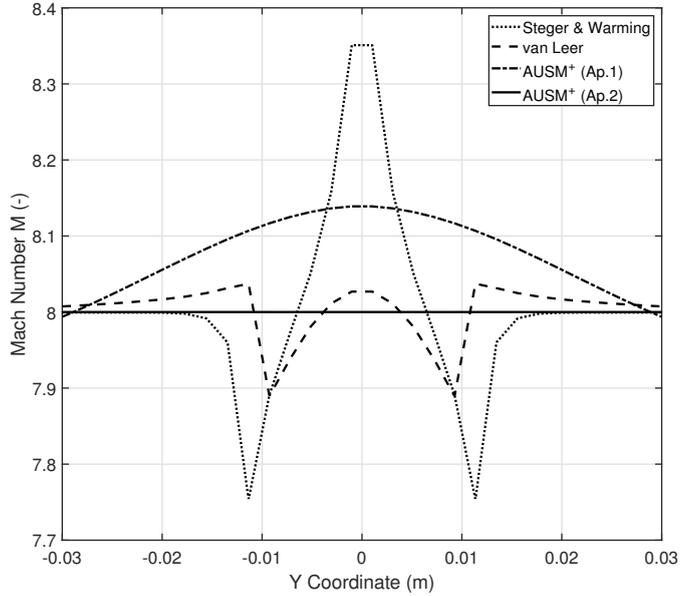

**Fig. 7** Calculated Mach number along a vertical line located in the upstream region of the shock, at $x = -0.07\,m$, for four different first-order upwind schemes. A constant freestream state of $M_\infty = 8.0$ is expected.

Mach number values can reach a difference of up to 4.4% in relation to the analytical freestream solution state, followed by 1.7% for the $\text{AUSM}^+_{Ap.1}$ and 1.4% for the van Leer scheme. Interestingly, the $\text{AUSM}^+_{Ap.2}$ scheme does not suffer from this problem. In locations far enough from the centerline of the domain, not pictured in Fig. 7, the freestream state is recovered by all first-order upwind schemes.

An interesting observation that can be made based on Fig. 7 is that even though the oscillations themselves are of pure numerical nature, their behavior still follows the expected relative performance between each one of the schemes. Both Steger and Warming and van Leer schemes, which are based on similar approaches for performing the flux vector splitting, produce oscillations of similar patterns. The latter scheme, however, produces oscillations of smaller magnitude when compared to the former one. This follows the trend already discussed in the literature [22], in which the van Leer scheme is capable of achieving better results than the Steger and Warming scheme for transonic and supersonic flows. On the other hand, while the oscillations produced by both Steger and Warming and van Leer schemes have well-defined edges, the ones produced by the $\text{AUSM}^+_{Ap.1}$ scheme are completely smooth. Furthermore, the simple modification made in the definition of the $\text{AUSM}^+_{Ap.2}$ scheme completely removed these oscillations, which are nonexistent up to machine precision.

As discussed by Pulliam [19], the correct capture of the freestream state when using discrete general curvilinear coordinates requires that the invariants of the domain transformation are satisfied even in discrete form. However, this is, in general, not the case. These invariants are partial differential equations themselves, which are equal to zero and primarily dependent on the metric terms. Inaccuracies and inconsistencies in the discrete treatment of the metric terms are the main reasons associated with non-zero values for these invariants. Therefore, the discrete metric terms are the major source of these oscillations. The $\text{AUSM}^+_{Ap.1}$ scheme, as employed here, utilizes a continuous well-behaved polynomial function to reconstruct the metric terms that accompany the pressure related components of the separated flux vectors [8]. Hence, this same continuous property is also seen in the behavior of the oscillations introduced by the $\text{AUSM}^+_{Ap.1}$ scheme, a contrast to the ones introduced by both the Steger and Warming and van Leer schemes. Since the metric terms are functions of the mesh geometry, it is expected that this problem becomes more apparent with the increase in mesh curvature. This leads to higher gradients in the metric terms, which increase the error associated with the invariants of the domain transformation.

In order to solve the oscillation problem, it was seen by the authors that by simply performing the freestream subtraction, a considerable reduction in the magnitude of the oscillations is achieved. For the Steger and Warming scheme, a reduction of at least 6 orders of magnitude is observed for the difference between the computed Mach number contours and the expected freestream value. Similar improvements in the quality of the solution were also observed for the van Leer and $\text{AUSM}^+_{Ap.1}$ schemes when performing the freestream subtraction. Differently from the $\text{AUSM}^+_{Ap.2}$ scheme, though, none of these three schemes were capable of achieving a fully oscillation-free solution. From a practical point of view, however, the obtained Mach number values behave as expected from the analytical solution for the freestream region. Therefore, for now on, all schemes analyzed here use the freestream subtraction.

Another problem, also seen in Fig. 5, is the deformation that the first-order van Leer scheme induces over the bow shock surface in the region that surrounds the symmetry line. Apparently, this deformation also has the same origin as the freestream oscillations, since it was observed that it completely disappears once the freestream subtraction is in place. Figure 8 shows the Mach number contours for the first-order van Leer scheme with freestream subtraction. One can clearly see that no oscillations are present in the freestream region of the flow, as well as no significant deformations exist in the shock surface.

For the second-order schemes, however, a different set of problems arises. First, both upwind schemes constructed with pure second-order operators, promptly diverge when performing the time march. When looking at the pseudo-transient solution, it is clear that the divergence comes from induced numerical oscillations that appear parallel to the surface of the shock. These oscillations quickly increase in magnitude, leading the solution to eventually diverge. This behavior can be attributed to the lack of sufficient artificial dissipation introduced by the schemes. One possible way of solving this problem is to simply explicitly add artificial dissipation to the discrete equations, such as the Pulliam nonlinear artificial dissipation model, defined in Eqs. (11) – (14). Another possible way of increasing the amount of artificial dissipation is to utilize first-order one-sided operators in the region that surrounds the shock wave. This can be achieved by using a flux limiter, as shown in the previous section. It was seen that either one of these approaches was able to fully stabilize the second-order upwind schemes, enabling them to reach a steady state regime. However, this was not possible without introducing some drawbacks.

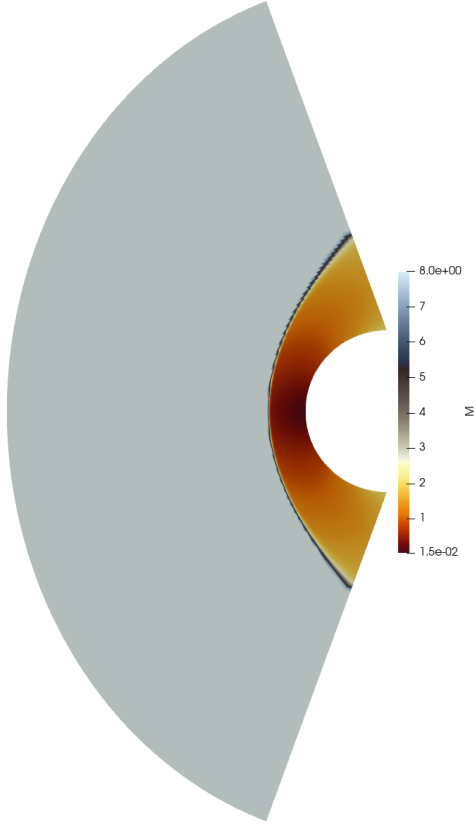

**Fig. 8** Calculated Mach number distribution for the inviscid blunt body flow using the first-order van Leer upwind scheme with freestream subtraction.

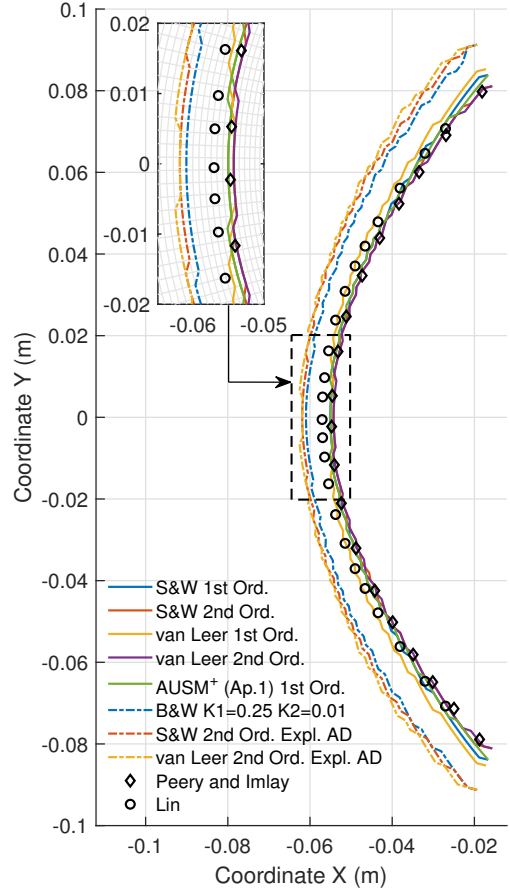

**Fig. 9** Approximate shock wave location obtained by the first and second-order schemes, with and without explicit artificial dissipation (AD). Results from Peery and Imlay [5] and Lin [6] are included for comparison.

Figure 9 shows the approximate shock wave location obtained by using the first and second-order schemes. This location is taken to be composed by the set of node coordinates, furthest away from the wall, in which the local Mach number has a 1% deviation from the freestream state. The results for the second-order upwind schemes are obtained by either using the minmod limiter, Eq. (28), or by explicitly adding the Pulliam artificial dissipation model, Eqs. (11) – (14). Numerical experiments performed by Peery and Imlay [5] and Lin [6], using modified versions of Roe's scheme to solve the full Navier-Stokes equations, are also added for comparison. Results for the $\text{AUSM}^+_{Ap.2}$ scheme are not added, since this scheme produces a virtually identical shock wave location as the $\text{AUSM}^+_{Ap.1}$ scheme with freestream subtraction. As clearly shown in the figure, there is significant aliasing in the shock surfaces obtained in the present work, for all considered schemes, when compared to the results from the other efforts. This is due to the misalignment between the shock surface and the mesh lines used in the present calculations. In the work of Peery and Imlay [5] and Lin [6], however, the mesh was built in such a way that a good alignment is obtained between the shock surface and the mesh lines. Furthermore, other numerical tools were also used in the referred work [5, 6] in order to smooth out this effect. Therefore, they do not suffer from this problem.

Another property, seen in Fig. 9, is that the solutions can be organized into two different groups based on the predicted shock wave location. The first group is composed by the methods that do not use explicit artificial dissipation, while the other one is composed by those that do use it. Notice that, when it comes to the location of the shockwave at the centerline of the domain, there is a significant gap between these two groups. It seems that the artificial dissipation model is moving the shock wave in the upstream direction. Although not shown here, when the original Beam and Warming artificial dissipation model [9] is used instead, a similar behavior is also displayed. Modifying the model constants does not change this behavior in any meaningful way. On the other hand, all upwind schemes, without the explicit addition of artificial dissipation, were able to achieve results that agree with the numerical data from the literature [5, 6], and position the shock wave slightly downstream as compared to the first group.

By observing the way in which the forward and backward operators are organized in Eq. (11), it is clear that the definition of the artificial dissipation model used here has a bias towards the backward direction. One could argue that this could be the reason for the observed numerical phenomenon. To verify this hypothesis, the forward and backward operators were switched in Eq. (11), resulting in Eq. (31),

$$
\begin{aligned}
D_\xi &= \Delta_\xi \left( \frac{\sigma_{i+1,j}^n}{J_{i+1,j}} + \frac{\sigma_{i,j}^n}{J_{i,j}} \right) \left( \epsilon_{i,j}^{(2)} \nabla_\xi \hat{Q}_{i,j}^n - \epsilon_{i,j}^{(4)} \nabla_\xi \Delta_\xi \nabla_\xi \hat{Q}_{i,j}^n \right), \\
D_\eta &= \Delta_\eta \left( \frac{\sigma_{i,j+1}^n}{J_{i,j+1}} + \frac{\sigma_{i,j}^n}{J_{i,j}} \right) \left( \epsilon_{i,j}^{(2)} \nabla_\eta \hat{Q}_{i,j}^n - \epsilon_{i,j}^{(4)} \nabla_\eta \Delta_\eta \nabla_\eta \hat{Q}_{i,j}^n \right),
\end{aligned}
\qquad (31)
$$

and the affected simulations were run once again. The observed solution has the shock wave translated approximately one node towards the body surface, which is a negligible amount, as seen from the visible mesh lines depicted in the inset of Fig. 9. Although capable of improving the results, the effect of this modification is so minimal that the order of the difference operators in the artificial dissipation model cannot be considered to be the root cause of the shock wave translation problem.

One can further observe that the differences among the calculated shock wave locations increase further away from the domain centerline, when comparing the solutions that are part of the same group. The different grouping patterns, though, are still well defined in these regions. When it comes to the quality of the solutions, it is clear that the schemes that do not use explicit artificial dissipation show good agreement with the data from Peery and Imlay [5]. This is especially true for the second-order upwind schemes, which are capable of reproducing the literature data very closely. The results from Lin [6], however, are located slightly upstream in relation to the first solution group. This difference in the results is constrained to a small region surrounding the centerline. This discrepancy can be due to inaccuracies introduced by the process of extracting the data from the original paper scan. Another possibility is that the dissipation model that is used in Lin's work, which has a term similar to the one used in the Pulliam model, is also slightly moving the shock upstream.

Although the use of a limiter to stabilize the second-order upwind schemes achieves better results than the explicit addition of artificial dissipation, a new disadvantage is introduced in the corresponding schemes. Now, the residue of the solution is unable to decrease 11 orders of magnitude, as obtained with the other schemes. Instead, a reduction of only 3 to 4 orders of magnitude is observed. This behavior is well documented in the literature [25, 26].

In Fig. 10, pressure coefficient ($C_p$) distributions along the cylinder surface are shown for the second-order schemes. The surface location is identified by the $\theta$ angle, which is measured from the center of the cylinder in relation to the centerline of the domain, in the counterclockwise direction. Upwind results are presented only for the calculations performed using a limiter, without introducing extra explicit artificial dissipation. Experimental results from Holden et al. [7] are also shown for comparison. As one can see, good agreement with experimental data is obtained for all schemes even though an inviscid formulation is used. Both second-order Steger and Warming and van Leer schemes obtained very similar results, with exception to the region between 40 and 60 deg. In this region, oscillations can be observed in the solution obtained with the Steger and Warming scheme. This region corresponds to the location of the sonic line and, therefore, it is susceptible to the formation of sonic glitches [22]. The results obtained with the Beam and Warming scheme are really close to the others for $\theta$ angles above 25 deg., even though its shock wave is located significantly upstream. On the other hand, the results obtained with the centered scheme yield a considerable deviation from the other results close to the centerline, which is where the subsonic flow pocket is located. Exactly at the centerline, for example, the calculated $C_p$ value is approximately 2.7% lower than the values obtained by the upwind schemes. Due to the dispersion present in the experimental data in this region, it is difficult to assess the relative performance of the schemes in relation to the quality of the $C_p$ values at low $\theta$ angles.

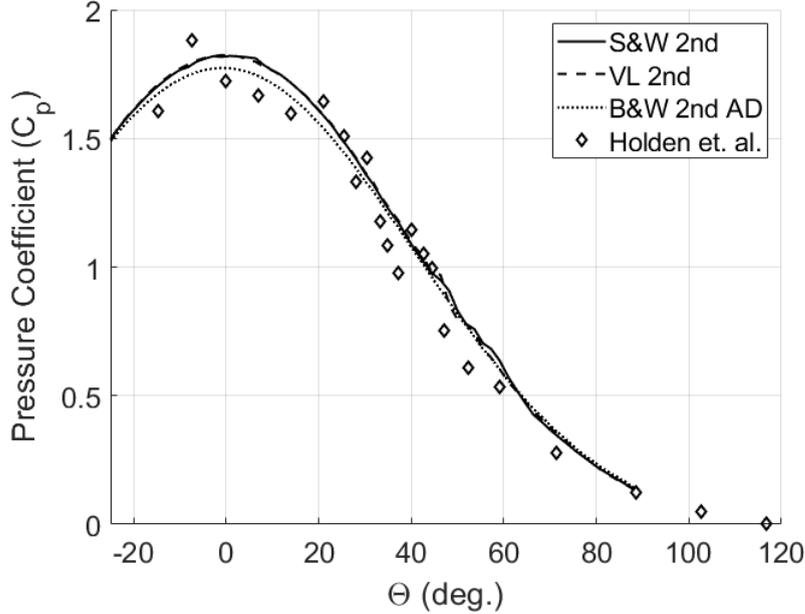

**Fig. 10** Pressure coefficient ($C_p$) distribution along the surface of the cylinder. The $\theta$ angle is measured from the center of the cylinder in relation to the centerline of the domain, in the counterclockwise direction. Experimental data from Holden et al. [7] is included for comparison.

# 6 Concluding Remarks

The correct definition of the supersonic inviscid flow that surrounds a circular blunt body can impose a significant challenge to multiple numerical schemes. In the present paper, this problem has been modeled using the Euler equations and, then, solved using five different numerical schemes written in the context of finite differences. Care must be exerted by the CFD user in order to avoid adding non-physical structures to the calculated properties. The authors have shown that, for the simple case considered, significant oscillations can arise in the freestream region of the solution under certain circumstances. These disturbances of numerical nature can also deform the overall shape of the shock wave, especially in the region close to the centerline of the domain. The presence of these structures is related to the metric terms of the domain transformation, which are functions of the mesh geometry itself. The authors observed that, for equations written in general curvilinear coordinates and discretized using finite differences, a simple practical fix to these problems is to perform a freestream subtraction over the flux terms of the original system of conservation laws. This technique reduces the amplitude of the numerical disturbances by at least 6 orders of magnitude while also allowing a better definition of the shock profile. Since no drawbacks are introduced to the numerical solution by performing this simple modification, its use is highly recommended. The second approach proposed in the present work to calculate the metric terms at the numerical interfaces, which was a necessity in the reinterpretation of the AUSM$^+$ scheme to a finite difference framework, does not develop these oscillations even when freestream subtraction is absent. Therefore, it is the preferred AUSM$^+$ form in a finite difference context, among the two considered here.

For the flow condition addressed in the present test case, pure second-order upwind schemes are unstable, as expected. The use of a limiter or the explicit addition of artificial dissipation are seen to be capable of stabilizing the solution process. Nevertheless, results obtained using explicit addition of artificial dissipation models translate the location of the shock in the upstream direction. No clear reason for this behavior has been found yet, but the authors are reporting the observed behavior in the present paper. Hence, for the second-order schemes analyzed, significantly better agreement with the reference numerical data is obtained by employing a flux-limiter formulation together with an upwind scheme. In summary, for the schemes in which such numerical spurious constructions are adequately controlled, or absent, good agreement has been found between the numerical results and the data available in the literature.

**Acknowledgments.** The authors wish to express their gratitude to the São Paulo Research Foundation, FAPESP, which has supported the present research under the Research Grants No. 2013/07375-0 and No. 2021/00147-8. The authors also gratefully acknowledge the support for the present research provided by Conselho Nacional de Desenvolvimento Científico e Tecnológico, CNPq, under the Research Grant No. 309985/2013-7.